# Andreev Interferometers in a Strong Radio-Frequency Field

C. Checkley, A. Iagallo, R. Shaikhaidarov, J. T. Nicholls, V. T. Petrashov [*)]
*Department of Physics, Royal Holloway, University of London,
Egham, Surrey TW20 0EX, United Kingdom*

**Abstract**

We experimentally study the influence of 1–40 GHz radiation on the resistance of normal *(N)* mesoscopic conductors coupled to superconducting *(S)* loops (Andreev interferometers). At low *RF* amplitudes we observe the usual *h/2e* superconducting-phase-periodic resistance oscillations as a function of applied magnetic flux. We find that the oscillations acquire a π-shift with increasing *RF* amplitude, and consistent with this result the resistance at fixed phase is an oscillating function of the *RF* amplitude. The results are explained qualitatively as a consequence of two processes. The first is the modulation of the phase difference between the N/S interfaces by the RF field, with the resistance adiabatically following the phase. The second process is the change in the electron temperature caused by the RF field. From the data the response time of the Andreev interferometer is estimated to be $\tau_f$ <40ps. However there are a number of experimental features which remain unexplained; these include the drastic difference in the behaviour of the resistance at $\varphi=\pi$ and $\varphi=0$ as a function of RF frequency and amplitude, and the existence of a "window of transparency" where heating effects are weak enough to allow for the π-shift. A microscopic theory describing the influence of RF radiation on Andreev interferometers is required.

**Introduction**

The well-established state-of-the-art read-out schemes for superconducting quantum systems and extremely sensitive magnetic flux-meters are mainly based on Superconducting Quantum Interference Devices (SQUIDs) consisting of superconducting loops interrupted by quantum tunnelling barriers, the Josephson junctions (see e.g [1] and references therein). SQUIDs have been exhaustively investigated for more than 45 years and have proved to be exceptionally efficient and are considered irreplaceable in a vast number of applications. However, recently discovered novel quantum systems such as artificial "atoms" [2], basic elements for quantum information technologies and quantum "meta-materials" [3] comprised of the artificial atoms, as well as very sensitive detectors of macroscopic quantum tunnelling of magnetization [4] require even more sensitive read-out devices with much lower "back action". This led to the further developments of SQUID-based techniques using their non-linear high frequency properties [5].

A promising alternative to a SQUID is an Andreev interferometer [6,7,8,9], which consists of a normal *(N)* mesoscopic conductor connected to superconductors *(S)*. The *N/S* interfaces play the role of mirrors reflecting electrons via an unusual mechanism first described by Andreev [10]. In Andreev reflection, an electron on the N side incident of the N/S interface creates a Cooper pair on the *S* side, as well as a hole that retraces the electron trajectory on the *N* side. There is a fundamental relationship between the macroscopic phase of the superconductors and the microscopic phase of the quasiparticles [11]: the hole gains an extra phase equal to the macroscopic phase of superconducting condensate χ, and correspondingly the electron acquires an extra phase −χ. Quantum interference of Andreev reflected quasiparticles results in high sensitivity of Andreev interferometers to the phase difference $\varphi=\chi_1-\chi_2$ between the N/S interfaces with a potential to achieve higher than state-of-the-art fidelity, sensitivity and read-out speed that are paramount to fundamental research and numerous practical applications of superconducting quantum systems. To achieve such challenging aims extensive investigations of Andreev interferometers on a scale similar to that of SQUIDs are in order.

In this paper we report on experimental study of Andreev interferometers driven by an RF field.

**Experiment**

The Andreev interferometers consisted of normal parts made of silver contacting superconducting loops made of aluminium shown in Fig. 1 (a). The interferometers were fabricated using the ''lift-off'' electron lithography technique. The first layer was 50 nm thick silver, and the second and the third layers were 20 nm thick insulating spacer ($Al_2O_3$) and 40 nm thick aluminum films, respectively. Before the deposition of aluminium an Ar ion beam was used to etch the $c$ and $d$ ends of the silver cross. Using the value of $\rho l=2.5\times 10^{-10}$ $\Omega\cdot cm^2$ for Ag [12], $\rho$ is the resistivity, $l$ is the elastic mean free path, we calculated the values of the diffusion coefficient of electrons $D=\left(\frac{1}{3}\right)v_F l \approx 400\ cm^2/s$ and the coherence length $\xi_N = \sqrt{\frac{\hbar D}{2\pi k_B T}} \approx 200\ nm$ at $T=1$ K. The area of superconducting loop was $42\ \mu m^2$. The area of the N/S interface was about 10,000 $nm^2$, and the distance between the N/S interfaces was 3000 nm. The precision of the alignment of different layers was better than 100 nm. The substrate was silicon covered by its native oxide. The RF field was created using an on-chip antenna. The diagram of experimental setup and the sample box are shown in Fig. 1 (b), (c). The sample box was partially filled with absorber to suppress RF resonances. The four terminal resistance $R$ between the points $a$ and $b$ was measured using lock-in techniques in the frequency range 100 Hz -100 kHz with the current ($I_{in}$) and voltage ($V_{out}$) probes as shown in Fig. 1 (b).

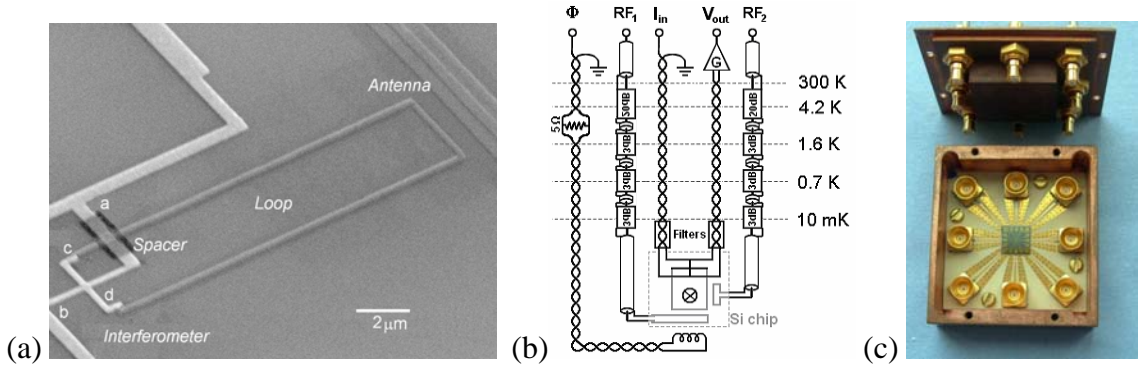

Fig. 1: (a) SEM photograph of a typical structure; the dimensions of interferometer and loop are given in the text. The resistance of the normal (silver) wire $ab$ with interfaces to superconducting aluminum loop at points $c$ and $d$ has been measured. A spacer of $Al_2O_3$ provides electrical insulation between the loop and the leads. (b) Schematic diagram of the measurement setup showing the filters and attenuators in the various lines. (c) Copper sample box where the sample chip is mounted on a PCB.

The resistance $R$ as a function of applied magnetic flux $\Phi$ followed the formula corresponding to a symmetric cross-like interferometer in the limit of weak proximity effect [8]

$$R(\varphi,T) = R_N - \delta R(T)(1+\cos\varphi),  \qquad (1)$$

where $R_N$ is the normal resistance of the mesoscopic conductor, $T$ is the temperature, and the superconducting phase difference $\varphi$ between $c$ and $d$ depends on the magnetic flux $\Phi$ through the loop via the relation $\varphi = 2\pi \frac{\Phi}{\Phi_0}$; $\Phi_0 = h/2e$. Figure 3 shows the dependence of the resistance $R$ on magnetic flux at different temperatures. The amplitude of the oscillations $\delta R$ was temperature dependent. At $\Phi=\Phi_0/2$ corresponding to $\varphi=\pi$ the resistance was close to its normal value. The resistance at other phases did not exceed the resistance at $\varphi=\pi$, as expected. The amplitude of oscillations decreases with increasing temperature. No "reentrance" [12,13] was observed, indicating that the Thouless energy in our samples was smaller than $k_B T$.

Figure 3 shows the resistance oscillations when subjected to 8.1 GHz radiation of different amplitudes. The magnitude of the oscillations decreases with increasing RF amplitude, and a $\pi$-shift is observed at high amplitudes. The dependence of the resistance on RF amplitude for four different values of $\varphi$ is shown in Fig. 4. The dependence shows oscillations superimposed on a monotonous decrease in the amplitude. Note that the resistance measured at $\varphi=\pi$ decreases substantially below the normal value for certain RF amplitudes. This fact is not accounted for by Eq. 1. A region where the flux-periodic oscillations show phase flips is indicated by an arrow.

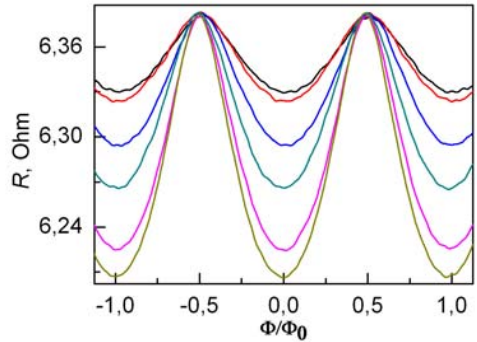

Fig. 2. The resistance $R$ measured as a function of applied magnetic flux $\Phi$ at mixing chamber temperatures of 50, 100, 200, 300, 500, and 600 mK.

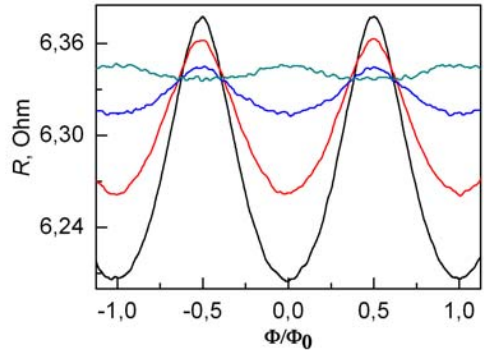

Fig. 3. The resistance $R$ as a function of applied magnetic flux $\Phi$, at different amplitudes of the 8.1 GHz radiation measured at 50 mK.

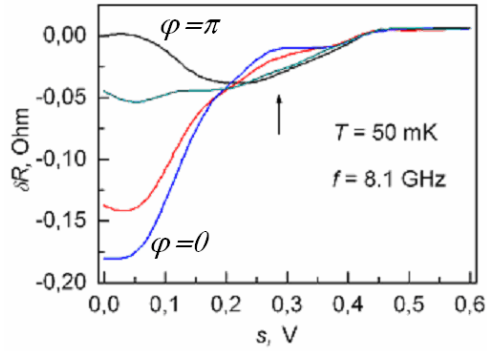

Fig. 4. The resistance *R* measured as a function of the amplitude of 8.1 GHz radiation. The different traces were measured at four values of constant flux, corresponding to the phase differences of *φ=0, π/3, 2π/3,* and *π*.

As shown in Fig. 5 the dependence of the resistance at a given flux on the RF amplitude was found to be strongly frequency-dependent. Indeed at 11.8 GHz (see Fig. 6) the oscillations as a function of RF amplitude are seen but there are no phase flips.

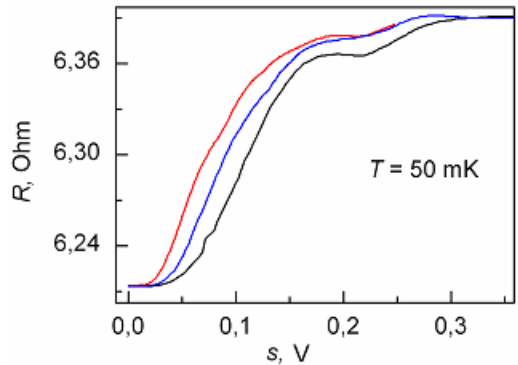

Fig. 5. The resistance *R* as a function of the RF amplitude at phase difference *φ=0* for different frequencies f=8 GHz, 8.2 GHz and 8.6 GHz.

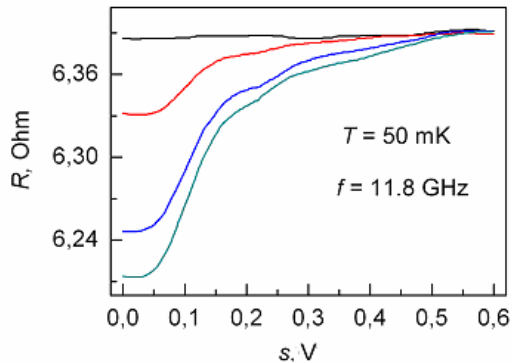

Fig. 6. The resistance *R* at the RF frequency f=11.8 GHz as a function of RF amplitude at different fluxes, corresponding to phase differences of *φ=0, π/3, 2π/3,* and *π*.

Figure 7 (a) shows the dependence of the measured resistance at $\varphi=\pi$ and $\varphi=0$ on frequency at a fixed amplitude of RF field. The dependence at $\varphi=\pi$ shows resonance-like dips with the positions and line shape strongly dependent on the RF amplitude (Fig. 7 (b). The measured resistance at $\varphi=0$ at fixed amplitude of RF vs. frequency on the contrary shows resonances (Fig. 7 (c) with a shape that is independent of RF amplitude. The dependence of resonance peak numbers, n, as a function of their frequency, $f_n$ was linear with a slope *df/dn* of about 1 GHz . Based on the time-domain measurements of the scattering parameters of our RF lines, we conclude that the resonances result from impedance mismatches in the measurement lines.

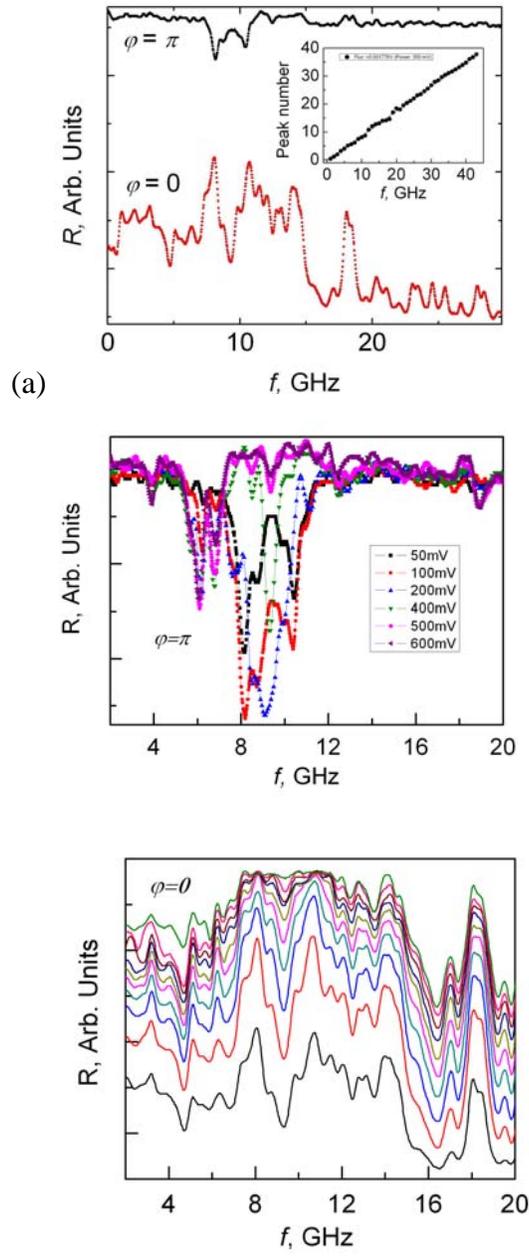

(a)

(b)

(c)

Fig. 7. (a) The resistance *R* as a function of the RF frequency for a fixed RF amplitude of 50 mV at $\varphi=\pi$ (upper curve) and $\varphi=0$ (bottom curve).. The inset shows the resonance peak number plotted as a function of frequency. (b) The resistance *R* at $\varphi=\pi$ as a function of the RF frequency at different RF

amplitudes from 50 mV to 600mV. (c) The resistance $R$ at $\varphi=0$ as a function of the RF frequency at different RF amplitudes from 50 mV to 600mV.

**Discussion**

The results shown in Figs. 3 and 4 can be explained by a combination of two processes. The first process is the modulation of the phase difference between the N/S interfaces, as a result of flux modulation by the RF field. The time-dependent phase can be written as

$$\tilde{\varphi} = \varphi + \mu \sin(2\pi t / \tau), \qquad (2)$$

where $\varphi$ is the phase across the interferometer due to the static magnetic field, and $\mu$ is the amplitude of modulation that is proportional to the amplitude of the RF field, $s$: $\mu = k \cdot s$, $k$ is the coefficient depending on transmission lines, coupling to the antenna etc. The value of $k$ is expected to be independent of $s$ for linear systems. The second process is the change in the electron distribution function caused by the RF field. We assume a model of heating with effective electron temperature $T^*=T+dT$. In general one should expect $dT$ to be a function of $s$, $f$ and $\varphi$, depending on the actual mechanism of interaction of electromagnetic field with electron-phonon system of Andreev interferometer. We assume a simple heating with $dT=\alpha s^2$, where $\alpha$ is a fitting parameter. In the adiabatic regime, when the response time of the interferometer $\tau_f$ is shorter than the period of the RF field $\tau = 1/f$, the instantaneous value of the resistance is

$$\tilde{R}(\varphi, T^*) = R_N - \delta R(T^*)(1 + \cos\tilde{\varphi}), \qquad (3)$$

where $T^* = T + as^2$ and $\varphi = 2\pi\dfrac{\Phi}{\Phi_0} + \mu\sin(2\pi t/\tau)$. It is assumed that the response time of the interferometer is of the order of the "time-of-flight" $\tau_f \approx L_{cd}^2/D$, and the resistance follows the RF-induced changes in the phase $\varphi$ adiabatically and $\omega\tau_f = 2\pi f \tau_f \ll 1$. In the limit of a weak proximity effect [8] the function $\delta R$ is phase independent and can be interpolated with the following function simulating non-monotonic dependence of the amplitude of oscillations due to "thermal" effect [13] and "re-entrance" [14] to the normal state at low temperatures

$$\delta R(T^*) = \delta R_0 \frac{T^*}{T_{Th}} \exp(-T^*/T_{Th}). \qquad (4)$$

The measured resistance can be written as

$$R(\varphi, T^*) = R_N - \delta R(T^*) \frac{1}{\tau} \int_{-\tau/2}^{\tau/2} \left(1 + \cos\left(\varphi + \mu\sin(2\pi t/\tau)\right)\right) dt. \qquad (5)$$

Simulations at $\varphi=0$ for different values of $\alpha$, describing the strength of heating effect, are shown in Fig. 8; the model describes qualitatively the features presented in Figs. 5 and 6. The integral in Eq. 5 yields a Bessel function, giving oscillatory behavior of the resistance as a function of RF amplitude. Similar oscillations have been observed[15] in SQUIDs.

Simulations for different values of magnetic flux at a fixed value of $\alpha$ are shown in Fig. 9, showing that the model can account qualitatively for the results shown in Fig. 4. The absence of the $\pi$-shifts in Fig. 6 can be simulated with $\alpha$ (which quantifies the strength of heating effect) depending on the phase $\varphi$. With $\alpha$ for $\varphi=\pi$ being sufficiently larger than for $\varphi=0$, the $\pi$-shifts disappear. The results presented in Fig. 7 can be accounted for by absorption of resonant RF photons in the Andreev interferometer.

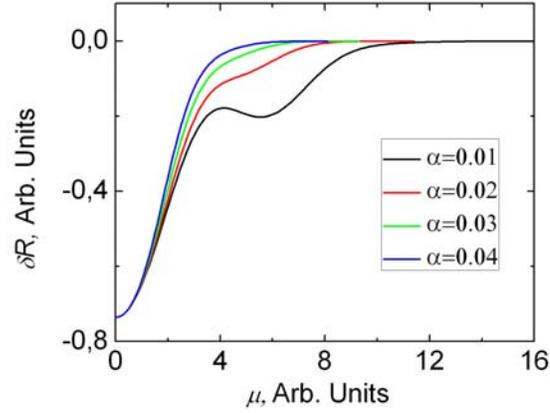

Fig. 8. Calculated resistance vs. RF amplitude at $\varphi=0$ for different values of the parameter $\alpha$ (see text).

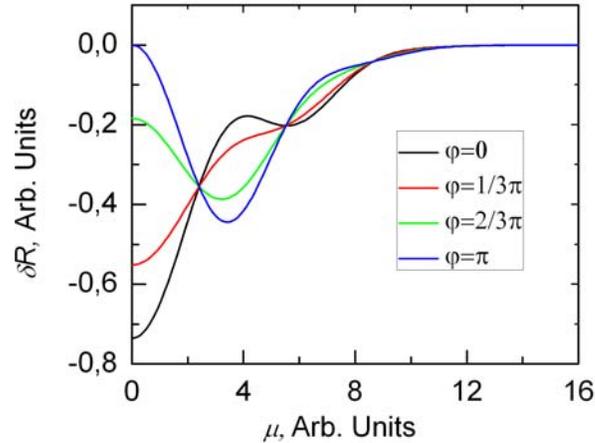

Fig. 9. Calculations of the resistance $R$ at a fixed RF frequency and parameter $\alpha$ as a function of RF amplitude at different fluxes, corresponding to phase differences of $\varphi=0, \pi/3, 2\pi/3, \pi$.

At high frequencies $\omega\tau_f = 2\pi f \tau_f > 1$ ($\hbar\omega > \varepsilon_{Th}$, where $\varepsilon_{Th} = h/\tau_f$ is the Thouless energy) the value of $k$ can be affected by the retardation effect. In this limit the adiabaticity breaks down and the resistance no longer follows the RF-induced changes, and so the "Bessel effect" is suppressed. This allows us to estimate the time of flight of the electrons in the interferometer. We observe oscillatory dependence of the resistance on the amplitude of RF field up to 13 GHz, which means that retardation effects are weak up to 26 GHz. The response time of the Andreev interferometer can be estimated to be $\tau_f <40$ps.

There are several experimental features which cannot be explained by our simple theory: (i) we observe oscillations as a function of RF amplitude which are not periodic. This implies that the coefficient *k*, which connects the amplitude of the RF modulation of the phase in Eq.2, is dependent on the amplitude; (ii) we cannot explain a "window of transparency" (Fig. 7 (b)) where heating effects are weak enough to allow for the "Bessel effect" that we believe can be the only explanation of the decrease in the resistance at $\varphi=\pi$ below its normal value; (iii) as yet we cannot explain the dramatic difference in the behaviour of the resistance at $\varphi=\pi$ and $\varphi=0$ as a function of RF frequency and amplitude, as shown in Figs. 7 (b) and (c); (iv) the time-of-flightestimated using the "static" diffusion coefficient from the dc resistivity is $\tau_f =200$ ps, which is larger than the "dynamic" value estimated from retardation effects.

**Conclusions**

The phase-periodic resistance oscillations in normal *(N)* mesoscopic conductors coupled to superconducting *(S)* loops (Andreev interferometers) driven by *RF* field acquire a *π*-shift with increasing amplitude. The resistance at fixed phase *φ* is an oscillating function of the *RF* amplitude, approaching normal resistance value at high amplitudes. The results can be explained as a combination of two processes. The first is the high frequency modulation of the phase by the RF field with the phase-periodic resistance adiabatically following the phase, resulting in the "Bessel effect". The second process is the heating by the RF field. There remain a number of unexplained experimental features; the high frequency properties of Andreev interferometers merit further experimental and theoretical investigation.

Acknowledgements. This work was supported by the Engineering and Physical Sciences Research Council (UK), Grant EP/E012469/1.

*) v.petrashov@rhul.ac.uk